\documentclass{kluwer}    

\newdisplay{guess}{Conjecture}

\begin{document}                                                                                   
\begin{article}
\begin{opening}         
\title{GRB: A Signature of Phase Transition to QGP?\thanks{ 
Supported by the National Natural Science Foundation of China under
Grant No.19803001,
by the Climbing project, and by the Youth Science Foundation of Peking
University.
            }} 
\author{R. X. \surname{Xu},
	Z. G. \surname{Dai}\footnote{
	Department of Astronomy, Nanjing University, Naijing 210093
	},
	B. H. \surname{Hong},
	G. J. \surname{Qiao}}  
\runningauthor{R. X. Xu, et al.}
\runningtitle{GRB: A Signature of Phase Transition to QGP?}
\institute{
	CAS-PKU joint Beijing Astrophysical Center
        and Department of Astronomy,
        Peking University, Beijing 100781, China
}
\date{}

\begin{abstract}
It is suggested that the inner energetic engine of Gamma ray burst
(GRB) may be the result of the transition of normal hadron to
quark-gluon plasma (QGP) in rapidly-rotating and spin-down
newborn neutron star. When such a nascent neutron star slows down
through dipole electromagnetic and quadruple gravitational radiation,
the increasing center density may reach the QCD transition density,
i.e., 5-10 nuclear density. Such kind of energy release from the phase
transition would be responsible for GRB and its possible beaming
effect. The relative dense gaseous environment of GRB location and
the iron line observed in the X-ray afterglow support this idea. Some
predictions in this model are given.
\end{abstract}

\end{opening}

Quark gluon plasma, supernova, and Gamma-ray burst are essential
puzzles in particle physics, stellar physics and high-energy
astrophysics, respectively. However, there might be an unusual
conjunction of them, from which the three problems originated. Here
we propose a possible way to settle the puzzles.

\vspace{4mm}
\noindent
{\bf Quark Gluon Plasma (QGP)}
From 1960s to 1990s in this century, one of the most important progresses
is the establishment of the standard model in particle physics. In
the model, there are four kinds of fundamental interactions between
elementary particles, namely, the gravitational interaction, the
electro-weak interaction, the strong interaction, and the Yukawa
interaction between Higgs particles. However, the knowledge of strong
interaction is much less than that of gravity or electro-weak
interaction. It is believed that the quantum chromodynamics (QCD)
developed in 1970s could be the successful dynamical theory of the
strong interaction, but the non-perturbation effect in QCD can hardly
be settled. According to lattice QCD, where discrete points represent
space-time, a new state of strong interaction matter, the so-called
quark-gluon plasma (QGP), appears when the temperature (up to $\sim
150-200$ MeV) or density (up to $5-10 \rho_0$, $\rho_0 \sim 3\times
10^{14}$g/cm$^3$ is the nuclear density) is large enough.

One of the QGP might be simply the lumps of up, down, and strange quarks
(also a few electrons for electric neutrality), which is known as strange
quark matter (SQM). Unfortunately, we can not tell whether SQM is the
lowest state of hadronic matter, but can only say that the energy of
strange matter is lower than that of matter composed by nucleus for
QCD parameters within rather wide range (Bodner 1971, Witten, 1984,
Farhi \& Jaffe 1984). Assuming SQM is absolutely stable, the most possible
QGP in nature may be strange stars (Alcock et al. 1986) which might be
the survivors in supernova explosions. The assumption is very {\it strong},
but {\it does not} be impossible.

While, in the terrestrial physics, it is the primary goal of
relativistic heavy-ion laboratory to search QGP (Muller 1995). Many
proposed QGP signatures have been put forward in theory and analyzed
in experimental data, but the conclusion about the discovery of QGP
are ambiguities. More likely, it is suggested in theory that 
the so-called strange hadron cluster (Schaffner et al. 1993) and
strangelet (Benvenuto \& Lugones 1995) may possibly exist in the nature, 
howerev, no experiment has affirmed or rejected the suggestion.

As will be suggested below, GRB might {\it also} be a signature of phase
transition from neutron matter to QGP composed of $u$, $d$, $s$, $e$
(i.e., SQM). If this can be confirmed by future astrophysical
observations, people can study the transition as well as the plasma
itself at a much longer timescale than those relevant to
terrestrial laboratory.

\vspace{4mm}
\noindent
{\bf Supernova Explosion (SNE)}~~~
In the 20th century, one of the prominence successes in astrophysics
is the develogment of the theory of stellar structure and evolution.
However, there are still some challenges in understanding stars' life,
such as the core-collapse supernova paradigm, which begins with the
collapse of the iron core of an massive star at the end of
its thermonuclear evolution (Bethe 1990, and references therein).
It is currently
believed that the prompt shock, which is the result of the inner core's
rebound after implosion has compressed the inner core to
supranuclear density, can not propagate directly outward and expel
the entire envelope, but stalls and turns into an accretion shock at
a radius of $100\sim 200$ km because of nuclear dissociation and neutrino
cooling. One way to rescue an explosion may be through the Wilson
mechanism (Wilson 1985, Wilson \& Bethe 1985), in which the neutrino
from the core can be absorbed by the material at $100\sim 200$ km and can
heat this material sufficiently to revive the shock to expel the
envelope. Generally, the simulations based on this mechanism give
rather low energy, 0.3-0.4 foe (1 foe $=10^{51}$ ergs), while the
observed energy in SN 1987A is at least 1.0 foe. Further, the supernova
simulations that do not incorporate fluid instabilities might fail
to explode (Bruenn 1993, Cooperstein 1993, Wilson \& Mayle 1993). Thus,
much of current research on the core collapse supernova mechanism is
focussed on the role of convection in the unstable regions that
develop below and above the neutrinosphere. Whereas, Mezzacapa et al.
(1998a, 1998b) found that the convection overturn and its associate
effects are not strong enough to revive the stalled prompt shock,
although the outward motion of the shock is enhanced.

Therefore, the key criterion for a successful explosion and its enough
energy may be the sufficient power of neutrino energy deposition behind
the stalled shock. The phase transitions from nuclear matter to 
two-flavor quark matter and from two-flavor quark matter to strange
quark matter (SQM) might be one possible way to increase the neutrino power
(Gentile et al. 1993, Dai et al. 1995). Because
each nucleon contributes about 30 MeV energy (Cheng et al. 1998), the
total energy $E_{\rm pt}$ of the phase transitions is about
$
E_{\rm pt} = M c^2 \times {30 \over 931}
\sim 5.8 \times 10^{52} {M\over M_{\odot}}
$
ergs, where $M$ is the mass of the inner core. Because the time scale
of phase transition is much smaller (below $10^{-7}$ s, see Dai et
al. 1995) than that of neutrino diffusion ($\sim 0.5$ s, see e.g.,
Bethe 1990), the neutrino luminosity $L$ caused by such conversion
is about $E_{\rm pt}/0.5 \sim 10^{53}$ erg/s. In Wilson's
computations, the typical value of $L$ is $5\times 10^{52}$ erg/s
without considering phase transition to SQM, hence, the total $L$
could be $15 \times 10^{52}$ erg/s ({\it three} times that of Wilson value).
Thus, the simulated explosion energy should be about $0.3-0.4$ foe
$\times 3$, which can explain the observed value from SN 1987A. The
residual strange star should be expected to be bare (Usov 1998) because
of high rate of mass ejection. These bare strange stars being chosen
as the counterpoints of pulsars have some advantages (Xu \& Qiao 1998,
Xu et al. 1999).

Another possible way favorable for a successful explosion might be
rapid rotation. A nascent neutron star can be rapidly rotating even
the initial precollapse core is not (Lai \& Shapiro 1995), for 
the radius of a typical white dwarf near the Chandrasekhar mass ($\sim
1.5 \times 10^3$ km) is about 10$^2$ times that of a neutron star
($\sim 10$ km). There may already be some indirect observational
evidences for an asymmetric core collapse (Lai \& Shapiro 1995 and
references therein), which may result in a rapidly rotating core.
Unfortunately, we have very little knowledge observationally and 
theoretically about how fast a nascent neutron star rotates. Further,
rapid rotation has not been treated extensively in the simulated
models. The existence of centrifugal force in a rotating core might
have at least two effects (Monchmeyer 1990): (1), to increase the
mass of inner core; (2), to enhance the prompt shock, thus favor a
successful explosion. The third effect, especially for the case of
rapid rotation, may be that the center density could be much lower
than the QCD transition density (Fig.~6.4 in Glendenning 1997) and therefore 
a {\it neutron star} (rather than a strange star) is left after
the explosion.

\vspace{4mm}
\noindent
{\bf Gamma Ray Burst (GRB)}~~~
Gamma-ray burst is one of the most challenges for physicists in this
and very possibly the next century. Although it is believed currently
that GRBs are at cosmological distances and the observed afterglow
can be well explained in the relativistic fireball models, the central
energetic engines responsible for these extraordinary events are
still poorly understood (e.g. Piran 1999, Meszaros 1999). Three main
classes of models are generally considered: the NS$^2$ merger model
(Eichler et al. 1989), the hypernova model (Paczynski 1998), and the
supernova model (Vietri \& Stella 1998).

The accumulating observations of afterglows have shown a close
relationship between GRBs and supernovae,
which are both the most energetic events in the universe.
Supernovae typically have explosion energy of 1 foe,
while the assumption of isotropic emission of GRB implies burst
energies in excess of 100 foe. There are evidences, which may be
suggestive of the hypernova or supernova models, that some detected
afterglows are in relatively dense gaseous environments which may
suggest that GRBs might be directly associated with star-forming
regions (Meszaros 1999, Piran 1999). Piro et al. (1999) reported the
possible detection (99.3\% of statistical significance) of
redshifted Fe ion line emission in the X-ray afterglow of GRB970508.
Having discussed possible mechanisms for the production of strong
iron line, Lazzati et al. (1999) found that a large amount of iron
might be in a compact region where a supernova exploded a few months
before the $\gamma$-ray burst. This result is favorable for the
supernova model.

We propose an alternative scenario of supernova model for GRBs. In the
Vietri-Stella one, a fast spinning neutron star, with a mass that
would be supercritical in the absence of rotation, is formed after
a supernova explosion. As it spins down, the neutron star inevitably
collapses to a Kerr black hole when its center density increases to
{\rm infinity}. Since SQM might be absolutely stable, the nascent
neutron star will undergo a transition of QGP to form a strange star
when the central density increases to the critical density for QCD
phase transition {\it before} collapsing to a Kerr black hole. Because
of the huge energy release in the transition ($\sim 10$ foe), mass
ejection is inevitable in the process, then it may be impossible
to form a Kerr black hole. In fact, this kind of phase transition has
been chosen as the energy source for GRB for a long time (e.g., Cheng
\& Dai, 1996, Ma \& Xie 1996). The obvious advantage in Cheng \& Dai's
model is that the baryon contamination is very small. The total burst
energy could be upto $\sim 10^3$ foe if the beaming effect is included
since a rapid spin neutron star or strange star is extremely non-
spherical-symmetric (see discussion below).

Gravitational radiation induces a generic instability in rapidly
rotating stars: a configuration with too much rotational energy will
radiate away its excess angular momentum until a stable configuration
is attained (Chandrasekhar 1970, Friedman \& Schutz 1975). A nascent
neutron star could be secularly unstable when the ratio $\beta=T/|W|$
of the kinetic to potential energy is greater than $\beta_{\rm
sec}\sim 0.14$, while, the star should be dynamically unstable when
$\beta>\beta_{\rm dyn}\sim 0.27$. Gravitational radiation will
destabilize some modes when $\beta>\beta_{\rm sec}$. For
$0<\beta-\beta_{\rm sec}\ll 1$ (i.e., secularly unstable but
dynamically stable), the growth time $\tau_{\rm GW}$ for the secular
instability is given by (Lai \& Shapiro 1995)
$$
\tau_{\rm GW} \sim 2\times 10^{-5} M_{1.4}^{-3} R_{10}^{4}
(\beta - \beta_{\rm sec})^{-5}~~{\rm s},
$$
where $M_{1.4}$ is the neutron star mass in unit of 1.4 $M_{\odot}$,
$R_{10}$ the neutron star radius in unit of 10 km. For typical neutron
stars, $\tau_{\rm GW}\sim 20$ s for $\beta\sim 0.20$, $\tau_{\rm
GW}\sim 7\times 10^4$ s for $\beta\sim 0.15$, and $\tau_{\rm GW}$
could be very long for a very small $\beta-\beta_{\rm sec}>0$ (Fig.~1
in Lai \& Shapiro 1995). For causal equation of state, Hasensel et al.
(1999) found the minimum periods of uniformly rotating neutron stars
can be as low as 0.288 ms.

If $\tau_{GW}>0.1$ yr, the usual magnetic dipole radiation might
become un-neglectable. The time scale $\tau_{\rm EM}$ for such
spin-down is (Vietri \& Stella 1998)
$
\tau_{\rm EM} \sim 0.1 {\rm yr}~ \omega_4^{-4} B_{13}^{-2}
$
for typical neutron stars, where $\omega$ and $B$ are the angular
frequency and the polar cap magnetic field of a neutron star,
respectively, and $\omega_4 = \omega/(10^4 {\rm s}^{-1})$, $B_{13} =
B/(10^{13} {\rm G})$. The magnetic field for nascent neutron stars
can be as strong as $3 \times 10^{15}$ G because of convection and
amplification of their magnetic field (Thompson \& Duncan 1993, Janka
1998). Therefore $\tau_{\rm EM}$ can range from hours to months.
Obviously, the time scale for substantial angular momentum loss can
be much shorter than the star's lifetime.

A newborn neutron star will be converted into a strange star when
spinning down enough that the central density increases to
5-10 $\rho_0$. For neutron stars, {\it only those very close to the
mass limit can rotate rapidly} because there the radius is the least
and the mass the greatest (i.e., the fine tune in mass, see Glendenning
1997), therefore, the time between supernova and conversion may be
smaller than $\tau_{\rm GW}$ or $\tau_{\rm EM}$.

When a conversion takes place, a large amount of energy ($E_{\rm pt}
\sim 58$ foe) is released. Also the rotational and the magnetic energy
would have important results in the burst of a nascent neutron star
into a strange star. In fact, dumping huge energy into a small spatial
volume in a short time inevitably leads to an opaque ``fireball''
(Meszaros \& Rees, 1993) because of large possibility for the
production of photon-photon pairs and electron-positron pairs.
Because of multiple or dipole magnetic field around a nascent neutron
star, the burst fireball might be significantly anisotropic; thus one
could have a sense of the GRB's {\it beaming effect},
which is very essentail for expalining the great burst luminosities.
The fireball could
preferably expand in weaker field regions with the open fields.
Furthermore, there might be a super-giant glitch (Ma \& Xie 1996)
accompanying the conversion since the equation of state of quark
matter is much softer than that of neutron star. This possible glitch
could results in strong electric field $E_{\rm gl}$ in the open field
line regions,
$$
E_{\rm gl} \sim R~~\nabla \cdot E_{\rm gl}
        \sim {RB\over 2\pi c} \delta \omega \\
        \sim 1.6 \times 10^{16} {\rm V~m^{-1}}
        {\delta \omega \over \omega}
        R_{10} B_{13} \omega_4,
$$
which can help charged leptons expanding in the region. Where $\delta
\omega$ is the change of $\omega$ when glitch, $R$ is the radius of
neutron star. $E_{\rm gl}\sim 10^{15}$ V/m if
${\delta \omega \over \omega}\sim 0.3$ (Ma \& Xie 1996).
In fact, Janka (1998) has found that such beaming effect might be
important for an anisotropic supernova explosion to produce kicks in
excess of 1000 km/s as long as an internal magnetic field could be
$\sim 10^{14}$ G.
In addition, the magnetic annihilation above a nascent neutron star
and strange star may also be effective for the variety of afterglows
(Dai \& Lu 1999).
In a conclusion, the energy sources responsible for GRB might be in
four parts: {\it phase transition}, {\it gravitational}, {\it
rotation}, and {\it magnetic energy}, although the magnetic energy
may be originated from rotation according to Thomson \& Duncan (1993).

\vspace{4mm}
\noindent
{\bf Conclusion}~~~
We have suggested a model that GRBs might be originated from the QCD
phase transition of rapidly-rotating and spin-down nascent neutron
stars to strange stars. The beaming effect, as well as the correlation
between GRBs and supernovae explosions, can be understood in the model.
Also the suggestion clarify the remain objects after later evolution
of massive stars: 1. White dwarfs (for less massive stars), 2. neutron
stars (for the case of highly rotating) or strange stars (for the case
that rotation is not important), 3. black holes (for very massive
stars). The neutron stars in case 2 should undertake a phase
transition to SQM, thus produce fireballs of BRBs.
Finally, the suggestion might be meaningful in laboratory physics.

{\it How can we distinguish this model from other models
observationally?}~~
According to Vietri-Stella model and our model, there is a supernova
before a GRB. Hence, we can monitor some sampled regions with supernovae.
There is possibility that a few of these samples can appear GRBs.
As the remnants after GRBs in Vietri-Stella model are black holes,
while those in our model are strange stars, we can differentiate these
models by further observations. As strange stars can act as pulsars
(Xu \& Qiao 1998, Xu et al. 1999),
we may find radio pulsars in GRB regions according to our model.


We are very grateful to Professors Y.R. Wang, T. Lu, Q.H. Peng,
S.Y. Pei, and J.M. Wang for their valuable discussions.

\end{article}

\begin{thebibliography}{}

\bibitem[\protect\citeauthoryear{}{}]{}
Alcock,C.,Farhi,E.,\&Olinto,A.,1986, 
	ApJ,310,261 
 
\bibitem[\protect\citeauthoryear{}{}]{}
Benvenuto, O.G., Lugones, G., 1995, 
	Phys. Rev. D51, 1989 
 
\bibitem[\protect\citeauthoryear{}{}]{}
Bethe, H.A., 1990, 
        Rev. Mod. Phys. 62, 801 
 
\bibitem[\protect\citeauthoryear{}{}]{} 
Bethe, H.A., \& Wilson, J.R., 1985 
        ApJ 295, 14 
 
\bibitem[\protect\citeauthoryear{}{}]{} 
Bodner, A.R., 1971, 
        Phys. Rev. D4, 160 
 
\bibitem[\protect\citeauthoryear{}{}]{} 
Bruenn, S.W., 1993, 
        in Nuclear Physics in the Universe, ed. M.W. Guidry \& 
        M.R. Strayer (Bristol: IOP), 31 
 
\bibitem[\protect\citeauthoryear{}{}]{} 
Chandrasekhar, S., 1970, 
        Phys. Rev. Lett. 24, 611 
 
\bibitem[\protect\citeauthoryear{}{}]{} 
Cheng, K.S., \& Dai, Z.G., 1996, 
        Phys. Rev. Lett. 77, 1210 
 
\bibitem[\protect\citeauthoryear{}{}]{} 
Cheng, K.S., Dai, Z.G., Wei, D.M., Lu, T., 1998, 
        Science 280, 407 
 
\bibitem[\protect\citeauthoryear{}{}]{} 
Cooperstein, J., 1993, 
        in Nuclear Physics in the Universe, ed. M.W. Guidry \& 
        M.R. Strayer (Bristol: IOP), 99 
 
\bibitem[\protect\citeauthoryear{}{}]{} 
Dai, Z.G., Peng, Q.H., \& Lu, T. 1995 
        ApJ, 440, 815 
 
\bibitem[\protect\citeauthoryear{}{}]{} 
Dai, Z.G., \& Lu, T. 1998, 
        Phys. Rev. Lett. 81, 4301 
 
\bibitem[\protect\citeauthoryear{}{}]{} 
Eichler, D., Livio, M., Piran, T., \&Shramm, D.N., 1989, 
        Nature, 340, 126 
 
\bibitem[\protect\citeauthoryear{}{}]{} 
Farhi, E., \& Jaffe, R.L., 1984 
        Phys. Rev. D30, 2379 
 
\bibitem[\protect\citeauthoryear{}{}]{} 
Friedman, J.L., \& Schutz, B.F., 1975, 
        ApJ 199, L157 
 
\bibitem[\protect\citeauthoryear{}{}]{} 
Gentile, N.A., et al., 1993, 
        ApJ, 414, 701 
 
\bibitem[\protect\citeauthoryear{}{}]{} 
Glendenning, N.K., 1997, 
        {\rm Compact Stars}, Springer-Verlag, New York 
 
\bibitem[\protect\citeauthoryear{}{}]{} 
Janka, H.-Th., 1998, 
        Proc. Of the 4th SFB 375 workshop on Astro-particle physics, 
        http://www.mpa-garching.mpg.de/SFB, No:SFB-375/296 09.01.98 

\bibitem[\protect\citeauthoryear{}{}]{}
Haensel, P., Lasota, J.P., Zdunik, J.L., 1997,
	A\&A, 344, 151 (astro-ph/9901118)
 
\bibitem[\protect\citeauthoryear{}{}]{} 
Lai, D., \& Shapiro, S.L., 1995, 
        ApJ 442, 259 
 
\bibitem[\protect\citeauthoryear{}{}]{} 
Lazzati, D., Campana, S., \& Ghisellini, G., 1999, 
        A\&A, in press (astro-ph/9906235) 
 
\bibitem[\protect\citeauthoryear{}{}]{} 
Ma, F., \& Xie, B., 1996, 
        ApJ 462, L63 
 
\bibitem[\protect\citeauthoryear{}{}]{} 
Meszaros, P., 1999, 
        The 19th Texas Symp. On Relativistic Astrophysics \& 
Cosmology (astro-ph/9904038) 
 
\bibitem[\protect\citeauthoryear{}{}]{} 
Meszaros, P. \& Rees, M.J., 1993, 
        ApJ 405, 278 
 
\bibitem[\protect\citeauthoryear{}{}]{} 
Mezzacapa, A., et al., 1998a, 
        ApJ 493, 848 
 
\bibitem[\protect\citeauthoryear{}{}]{} 
Mezzacapa A., et al., 1998b, 
        ApJ 495, 911 
 
\bibitem[\protect\citeauthoryear{}{}]{} 
Monchmeyer, R., 1990, 
        Ph.D thesis, Tech. Univ. Munchen 
 
\bibitem[\protect\citeauthoryear{}{}]{} 
Muller, B., 1995, 
        Rep. Prog. Phys. 58, 611 
 
\bibitem[\protect\citeauthoryear{}{}]{} 
Paczynski, B., 1998, 
        ApJ, 494, L45 
 
\bibitem[\protect\citeauthoryear{}{}]{} 
Piran, T. 1999, 
        Phys. Rep. in press (astro-ph/9904038) 
 
\bibitem[\protect\citeauthoryear{}{}]{} 
Piro, L., et al., 1999, 
        A\&A, in press 
 
\bibitem[\protect\citeauthoryear{}{}]{} 
Schaffner, J. et al. 1993, 
Phys. Rev. Lett. 71, 1328 
 
\bibitem[\protect\citeauthoryear{}{}]{} 
Thompson, C., \& Duncan, R.C., 1993, 
        ApJ 408, 194 
 
\bibitem[\protect\citeauthoryear{}{}]{} 
Usov, V.V., 1998, 
        Phys. Rev. Lett. 80, 230 (astro-ph/9712304) 
 
\bibitem[\protect\citeauthoryear{}{}]{} 
Vietri, M., \& Stella, L., 1998, 
        ApJ, 507, L45 
 
\bibitem[\protect\citeauthoryear{}{}]{} 
Wilson, J.R., 1985, 
        in {\it Numerical Astrophysics}, edited by J.M. Centrella, 
        et al. (Jones \& Bartlett, Boston), p.422 
 
\bibitem[\protect\citeauthoryear{}{}]{} 
Wilson, J.R., \& Mayle, R.W., 1993, 
        Phys. Rep., 227, 97 
 
\bibitem[\protect\citeauthoryear{}{}]{}
Witten, E., 1984, 
        Phy. Rev. D30, 272 
 
\bibitem[\protect\citeauthoryear{}{}]{}
Xu, R.X., \& Qiao, G.J., 1998, 
        Chin. Phys. Lett., 15, 934 (astro-ph/9811197) 
 
\bibitem[\protect\citeauthoryear{}{}]{}
Xu, R.X., Qiao, G.J., \& Zhang, B., 1999 
        ApJ, 522, L109

\end{thebibliography}
\end{document}